# Optical centroid orbiting metrology

Liang Fang[*], Jinman Chen, Qinjun Chen, and Chujun Zhao[*]

School of Physics and Electronics, Hunan University, Changsha 410082, China.

*Corresponding author: liangfang@hnu.edu.cn, cjzhao@hnu.edu.cn

## Abstract

Optical interferometry has dramatically advanced the development of modern science and technology. Here we introduce an interesting centroid evolution phenomenon of orbital angular momentum (OAM) interference fields with broken rotational symmetry, and establish a novel interferometric paradigm by fully exploiting centroid orbiting information. The centroid positions and their geometric trajectories can provide more detectable information in a two-dimensional plane to sense the interferometric perturbations, compared with the conventional interferometry. We first investigate centroid orbital evolution under the inclined angle perturbation that allows for ultra-sensitive angle distinguishment with arc-second resolution. We also show centroid ellipse evolution under spatial phase perturbation that enables geometric characterization of arbitrary OAM superpositions on modal Poincaré spheres. Furthermore, based on the angle subdivision of centroid orbiting, we demonstrate the environmentally robust nanoscale displacement measurement with polarization synchronous detection, and particularly the high-resolution, fast, and large-range linear movement monitoring using commercial four-quadrant photodetectors. This novel centroid orbiting interferometry may open new opportunities to advance metrological technologies beyond the conventional interferometers.

## Introduction

Metrology is indispensable for humans to sense the environment and even know the universe. Optical interferometric metrology benefits fast, noninvasive, and high-precision performance, and thus has dramatically made progress in modern science and technology[1,2]. Particularly, it has always played a key role in the fundamental researches about the physical universe. In 1887, Michelson-Morley experiment using the interferometer attempted to confirm the existence of a fixed frame of reference for wave propagation, while the negative result contributed to the special theory of relativity[3]. Nowadays, the basic configureuration of the Michelson interferometer continues to be on a mission to detect gravitational wave by the laser interferometer gravitational-wave observatory (LIGO)[4]. Besides, myriad categories of interferometers, as powerful metrological tools, have sprung up and broadly applied to the realms of mechanical engineering, quantum physics, material science, semiconductor industry, etc[5-11]. For the existing optical measuring instruments, the directly detectable physical quantities include optical intensity, frequency (wavelength), and polarization degrees of freedom (DoFs). The most common methods to extract interference information are to analyze the intensity variation of interference fringes or Newton's rings caused by optical path difference (OPD) changes (Figures 1a and 1b).

Beyond the well-known fundamental Gaussian beams with plane phase distributions, higher-order Laguerre-Gaussian (LG) or vortex beams with spatially helical phase carrying orbital angular momentum (OAM) have attracted a great deal of attention in optics community[12,13]. So far, optical beams have developed into a big family of structured light by tailoring the diverse electromagnetic dimensionalities, such as amplitude, phase, polarization, and even spin in a two- or three-dimensional (2D or 3D) space[14-21]. These structured fields manifest as many unique characteristics and have found broad applications in super-resolution microscopy, optical metrology and tweezers, quantum information processing, even classical and quantum communication systems, etc[22-34]. The interference or superposition fields between structured beams with different mode orders can shape composite fields into various spatial fringes or patterns, which provides a new detectable dimensionality for multi-dimensional or high-precision optical metrology[35-42].

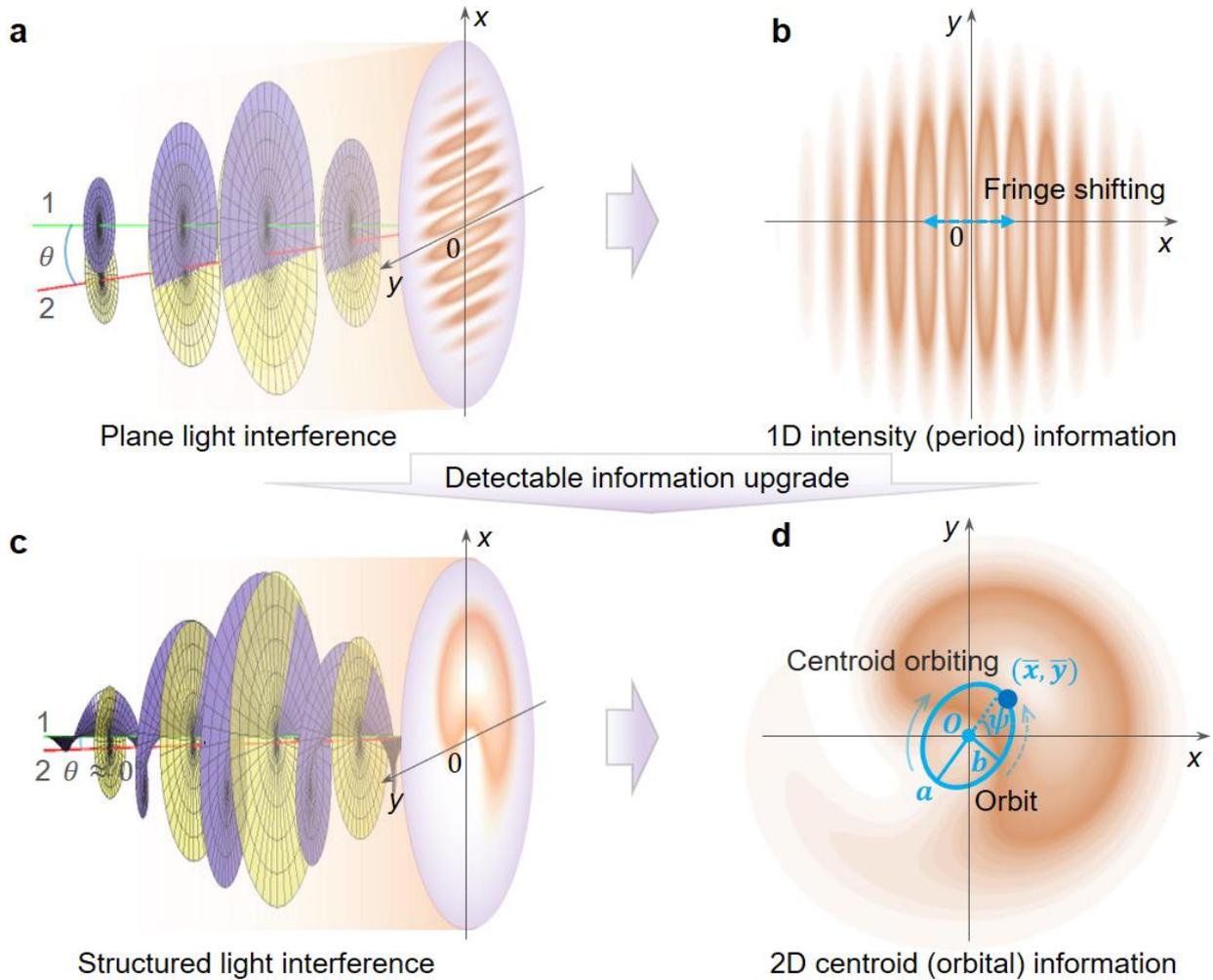

**Figure 1. Centroid orbiting provides more detectable DoFs than fringe shifting to sense interferometric perturbations.** (a) Conventional interference fringes produced by plane light beams ($\ell_1=0$, $\ell_2 = 0$) just carry (b) single (1D) intensity or period information. (c) Centroids as arithmetic mean positions of globally interference field of two LG beams ($\ell_1=1$, $\ell_2 = 0$) carry (d) diverse (2D) positional and orbital information. The upgraded detectable information contains centroid positions ($\bar{x}, \bar{y}$), orbiting velocities and directions, as well as the orbital parameters in terms of elliptical center (O), ellipticity ($e = b/a$), orientation angle ($\psi$) of major axis, and swept area ($S = \pi ab$).

Centroid (or center of mass) is a useful parameter for characterizing the arithmetic mean of all points weighted by the local density or specific weight in mathematics and physics. It is well known that for a pattern (object) with regular geometric shape and homogeneous density, its centroid always shares the same positions with the geometric center, while for this with irregular shape (broken rotational symmetry), the centroid is off center. Inspired by this, here we investigate the optical centroids of structured interference patterns of the commonly used LG beams, and unveil its evolution rules under various interferometric perturbations. Interestingly, the centroids can be unfolded to a specific orbit or trajectory by breaking rotational symmetry of the nearly coaxial interference patterns using structured light beams with non-uniform azimuthal phase distributions (Figure 1c). Note that if the light beams themselves feature non-symmetrical amplitude change in time and space, probably referring to some specific spatiotemporal beams, they may also form specific centriod trajectories. In this paer, we just focus on the phase-dependent centroid evolution in the frame of structured light interference.

Here we discover that the centroid coordinates inherently carry more detectable information than the interference fringes to sense interferometric perturbations, which contains diverse centroid orbiting information, especially the elliptic parameters in a 2D plane (Figure 1d), as well as the 3D orbiting directions (Figures 2b and 2c). This centroid information can be regarded as a new detectable quantity (or DoF) that may open new opportunities to advance optical metrological technologies. Here we organize this paper with four parts. In the first part, we present the physical origin and mathematical derivation of centroid orbiting based on the rotation of structured interference patterns. In the following three parts, we experimentally demonstrate the characteristics of centroid orbiting phenomenon and its potential applications. The first two experiments show centroid orbital evolutions under two typical perturbations, such as the tilt interference and OAM superpositions. The last one exhibits metrological advantages in the linear measurement by monitoring centroid orbiting angles on its circle orbit.

## Results

### Centroid orbital derivation and equations

For two-dimensionally spatial interference patterns, we first investigate the locations of their centroids. The centroid coordinates ($\bar{x}, \bar{y}$) of the interference patterns can be formulated by $\bar{x} = \iint r^2 \cos\phi \, I(r,\phi) dr d\phi / P$, and $\bar{y} = \iint r^2 \sin\phi \, I(r,\phi) dr d\phi / P$, where $P = \iint r I(r,\phi) dr d\phi$ is the total power, $(r, \phi)$ denotes the polar coordinate. $I(r,\phi) = |E_1|^2 + |E_2|^2 + 2|E_1 E_2|\cos[\Delta\ell\phi + kr\sin\theta \cos(\phi + \alpha) + k\Delta z + \Delta\xi]$ is the distribution function of interference patterns, where $E_1(r)$ and $E_2(r)$ represent the radial-dependent distributions of two interference fields, respectively, $k = 2\pi/\lambda$ is the wave number of light with the wavelength $\lambda$. $\Delta\ell = \ell_2 - \ell_1$, $\ell_1$ and $\ell_2$ are the topological charge numbers (TCNs) of LG beams, $\Delta z = z_2 - z_1$ is the OPD, $\Delta\xi$ is the phase differences for divergence functions, Gouy phases and initial phase between these two beams (see Supplementary Materials). Here we consider that one beam has an inclined (or polar) angle $\theta$ relative to the other and meanwhile has the azimuthal angle $\alpha$ around it, as shown in Figure. S1a.

In the nearly coaxial interference case ($\theta \to 0°$), following the orthogonality of trigonometric functions, the optical centroids can be unfolded to a specific orbit in the condition of $\Delta\ell = \pm 1$ that mathematically couple with the polar coordinates (see Supplementary Materials). In the low-order Taylor's approximation for the interference fields, the centroid orbital equations under the inclined angle perturbation ($\theta$ and $\alpha$) can be derived as,

$$\bar{x} \approx a \cos\vartheta \mp a^2 k\theta \sin\alpha, \tag{1}$$

$$\bar{y} \approx \mp a \sin\vartheta \mp a^2 k\theta \cos\alpha, \tag{2}$$

where $a = \zeta_1/\zeta_0$ is the radius of a perfect circular centroid orbit under completely coaxial interference between two beams ($\theta = 0°$), $\zeta_0 = \int(|E_1|^2 + |E_2|^2)r dr$ and $\zeta_1 = \int |E_1 E_2| r^2 dr$. This circular orbit of interference patterns essentially originates from helical phase in higher-order LG beams, directly associated with the skewed Poynting vector around optical axis[12,43]. $\vartheta = \varphi(t) + \Delta\eta$ determines the azimuthal positions (or phase) of centroids, where $\varphi(t)$ is the dynamical phase perturbation. It may result from OPD change or phase modulation for one light path, for example, introduced by normal reflection via a triangular prism (or mirror) with the shifting velocity of $v/2$, giving $\varphi(t) = k\Delta z(t) = kvt$, as shown in Figure 2(a). $\Delta\eta$ is the propagation-dependent Gouy phase and initial phase difference between the two interference beams. Here we assume that there are no divergences or nearly the same divergent degrees between them (see Supplementary Materials). The signs of '−' and '+' depend upon the TCN differences of $\Delta\ell = +1$ and $-1$, respectively, determining different orbiting directions (Figures 2b and 2c). In practice, these centroid orbiting directions can be distinguished via the signal advance or delay between *x* and *y* coordinate variations of centroids when projected on one time-domain window, as shown in Figures 2(b) and 2(c), respectively.

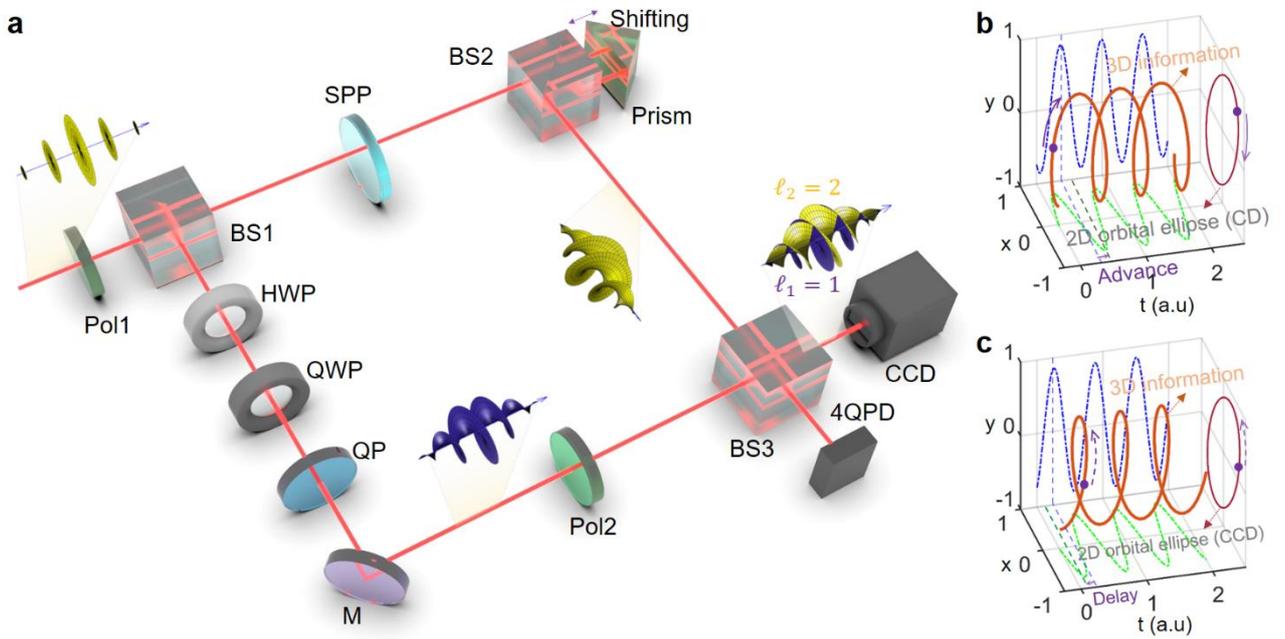

**Figure 2 Schematic of acquiring centroid geometric information.** (**a**) Experimental setup of extracting centroid from interference patterns using two LG beams with TCN differences of $\Delta\ell = +1$ or $-1$. Pol.: polarizer; BS: beam splitter; HWP: half wave plate; QWP: quarter wave plate; QP: Q plate; SPP: spiral phase plate; M: mirror; 4QPD: four-quadrant position-sensing photodetector; CCD: charge coupled device. The light beam with OAM of $\ell_1 = 1$ can be produced by QP in the down path, and the beam in the up path can be loaded with OAM of $\ell_2 = 2$ or 0 by using or removing SPP, corresponding to $\Delta\ell = +1$ and $-1$, respectively. The perfect circular centroid orbits easily evolve into elliptical orbits under various perturbations. (**b**) 3D centroid orbiting with clockwise direction (CD) produces the signal advance of the *x*-coordinates (green curve) relative to the *y*-coordinates (blue curve). (**c**) 3D centroid orbiting with counter-clockwise direction (CCD) produces the signal delay between them. Both 3D trajectories can be projected onto the *x-y* plane, forming the elliptical orbits.

**Centroid orbital shifting under inclined angle perturbation**
The second term in Equations (1) and (2) is responsible for the linear shift of centroid orbit away from the original center due to the inclined perturbation, which can be approximately given as

$$\Delta O \approx \mp a^2 k\theta = \mp \frac{2}{\lambda} S\theta, \quad (3)$$

where $S = \pi a^2$ is the swept area by the unfolded centroids around the axis. Remarkably, it shows that the centroid orbital area ($S$) enables amplification of the interferometric sensitivity to the inclined angles via the orbital shift ($\Delta O$). Note that these orbital equations are just satisfied in the condition of $k\theta \ll 1/a$. The completely low-order approximate equations, the orbital equations and their elliptic parameters in the higher-order Taylor's approximation are presented (see Supplementary Materials). Moreover, we numerically simulated the detailed evolution of centroid orbital shifts, ellipticity, and the orientation angles of the major axes over a large range of inclined angles (see Figure S3). The interference patterns trend to be the familiar fork-patterns when further enlarging the inclined angle. In this case, the centroid ellipses would become smaller and smaller until their centers approach to optical axis. In addition, as new detectable parameters, the centroids actually can move to form arbitrary trajectories with various orbital parameters in a 2D plane. For example, the centers of centroid ellipses are off and around optical axis when changing the azimuthal angle $\alpha$ (see Figure S4). In this case, the shifting directivities of ellipse centers are always parallel or antiparallel to the normal vector of the plane where the two axial lines ($z$- and $z'$-axis) lie, depending upon the signs of $\Delta \ell$, as shown in Figure S1a.

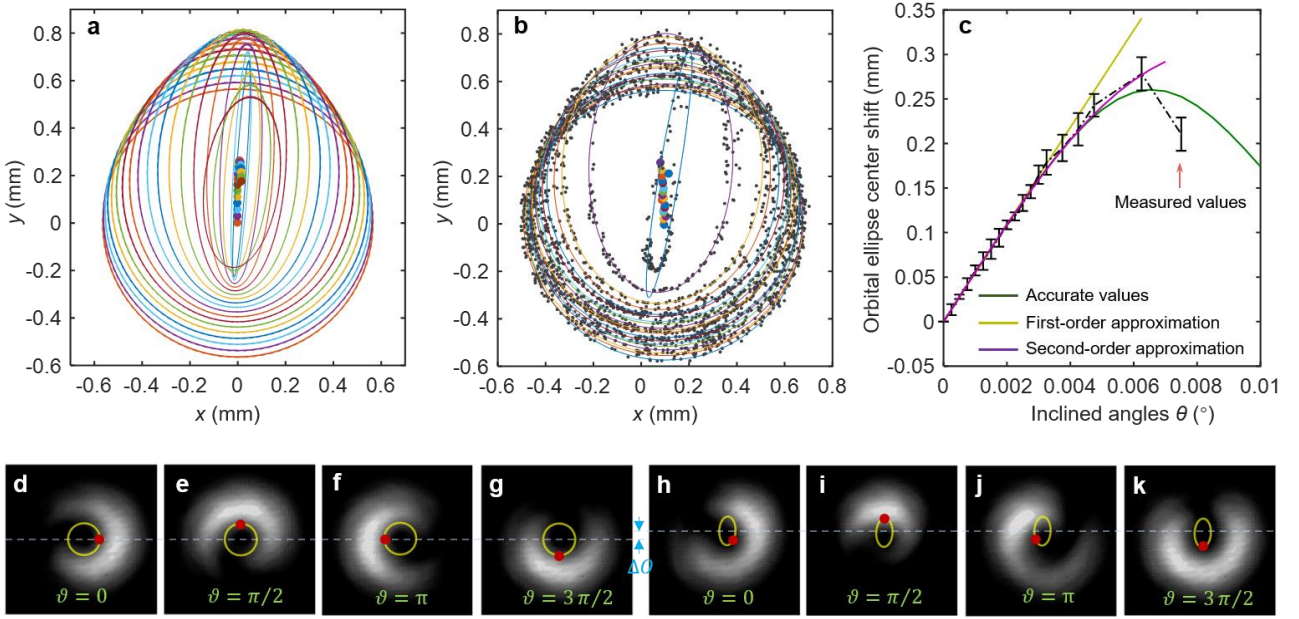

**Figure 3 Centroid orbital evolution and geometric center shifting under various inclined angles.** (**a**) The calculated results of centroid orbital shifting upwards and then downwards when the inclined angles $\theta$ change from $0°$ to $0.01°$ with a step of $0.0005°$ and the azimuthal angle is fixed as $\alpha = 0°$. (**b**) Measured results of centroid orbital shifting upwards and then downwards under successively varied inclined angles $\theta$ from $0°$ to $0.00325°$ with a step of $0.00025°$, and then $0.00375°$, $0.00425°$, $0.00475°$, $0.00625°$, and $0.0075°$, the azimuthal angle is $\alpha = 0°$. The dark dots denote the centroid positions extracted from the experimentally collected pictures via centroid algorithm, the lines represent the recovered ellipses via fitting algorithm, and the colourful dots indicate the center positions of the elliptical orbits. (**c**) The accurate, approximate, and measured results of the orbital center shift versus the inclined angles. Measured interference patterns under varied phase $\vartheta$ when the relative inclined angles

of two interference beams are (**d**)-(**g**) $\theta = 0°$ and (**h**)-(**k**) $\theta = 0.00625°$, respectively. The hot dots denote centroid positions and the yellow circles (or ellipses) indicate the orbits where the centroids locate.

We demonstrate the experimental results of ultra-sensitivity of centroid orbital shifts to the relative inclined angles ($\theta$) (Figure S1a). For convenience, the experiment was performed under the zero azimuthal angle ($\alpha \approx 0°$) of the probing beam relative to the reference beam. In the experiment, a perfect Gaussian beam with circular shape should be obtained, and the tiny inclined angles were generated by a six-dimensional alignment jig (see Materials and Methods). The probing and reference beams were loaded with the TCNs of $\ell_1 = 1$ and $\ell_2 = 2$ in order to get a bigger orbital size to sense the inclined angles (see the experimental setup and details in Figure S5). The dynamic interference patterns were produced by manually changing the OPDs, instead of shifting the rectangular prism in Figure 2a, and collected by a high-speed camera (about 160 fps). The centroid coordinates and their orbital ellipses were calculated via algorithm from the collected interference pictures (see Materials and Methods). We present the numerical and experimental results of the angle-dependent centroid orbits in Figures 3a and 3b, respectively. The measured values of centroid orbital shifts are plotted in Figure 3c, where also contain the theoretically predicted curve and the first-order and second-order approximate curves ($\Delta O$) for comparison. The measured orbital shifts are almost in line with the predicted and approximate curves, especially in the low-order linear approximation region (from $0°$ to $0.003°$). It is noteworthy that the centroid ellipses have a maximum shift (about 0.25 mm) when the inclined angle is $\theta_m \approx 0.0065°$ in the optical scene experimentally and numerically executed here. This special polar angle can be roughly derived from the centroid equations in the higher-order Taylor's approximation (see Supplementary Materials). In Figures 3d-3k, for comparison, we show the measured interference patterns with centroid positions and orbits under the changed phase $\vartheta$ when the inclined angles are $\theta = 0°$ and $0.00625°$. We also provide the numerical results (Figure S2) and the videos showing dynamical centroid orbiting (Videos S1 to S4). It is worth mentioning that generally the measurement based on the detection of centroid orbital information has the capacity of resistance to the environment disturbance. Because the environment disturbance mainly changes the centroid orbiting positions, but almost has no influence on the orbital parameters.

**Centroid ellipse evolution under OAM superposition perturbation**
The geometric evolution of centroid orbital ellipses can be produced by the perturbations not only arising from the interferometric environment, for example the tiny inclined angles, but also the OAM superpositions within structured light beams. In general, the orbital ellipses refer to the parameters in terms of ellipticity, orientation angle of major axis, swept area, and the orbiting directions. These may be exploited into a new metrological method to sense or quantify the spatial information of structured light beams. Based on this centroid ellipse evolution, here we demonstrate the geometric characterization of arbitrary superposition states of LG beams with opposite OAM (eigenbases) on modal Poincaré spheres[44]. A similar scheme has been reported about the usage of the centroid elliptical trajectories via mode transformation to measure geometric phase of structured Gaussian beams[45]. In our scheme here, the centroid ellipses are generated by coaxial interference between the reference and targeted beams both with opposite OAM components ($\pm\ell_1$ and $\pm\ell_2$, but limited by $|\ell_1| - |\ell_2| = \pm 1$), of which the equations can be deduced as,

$$\bar{x} = g[m_1\cos\varphi + m_2\cos(\varphi + \Delta\beta)], \qquad (4)$$

$$\bar{y} = g[-m_1\sin\varphi + m_2\sin(\varphi + \Delta\beta)], \tag{5}$$

where $g = \sqrt{2}\int|E_1 E_2|r^2\,dr / 2\int(|E_1|^2 + |E_2|^2)r\,dr$, $m_1$ and $m_2$ are the weight coefficients of opposite OAM components within the targeted LG beams, normalized as $m_1^2 + m_2^2 = 1$. $\Delta\beta$ is related with subtraction between the phase differences among OAM components for both reference and targeted LG beams, respectively (see Supplementary Materials). Here we set that the reference beam has the same weight coefficients. Note that the dynamical phase perturbation $\varphi$ also can be introduced by the OPD given as $\varphi(t) = k\Delta z(t) = kvt$ via a mirror with the moving velocity of $v/2$, similar to Equations (1) and (2).

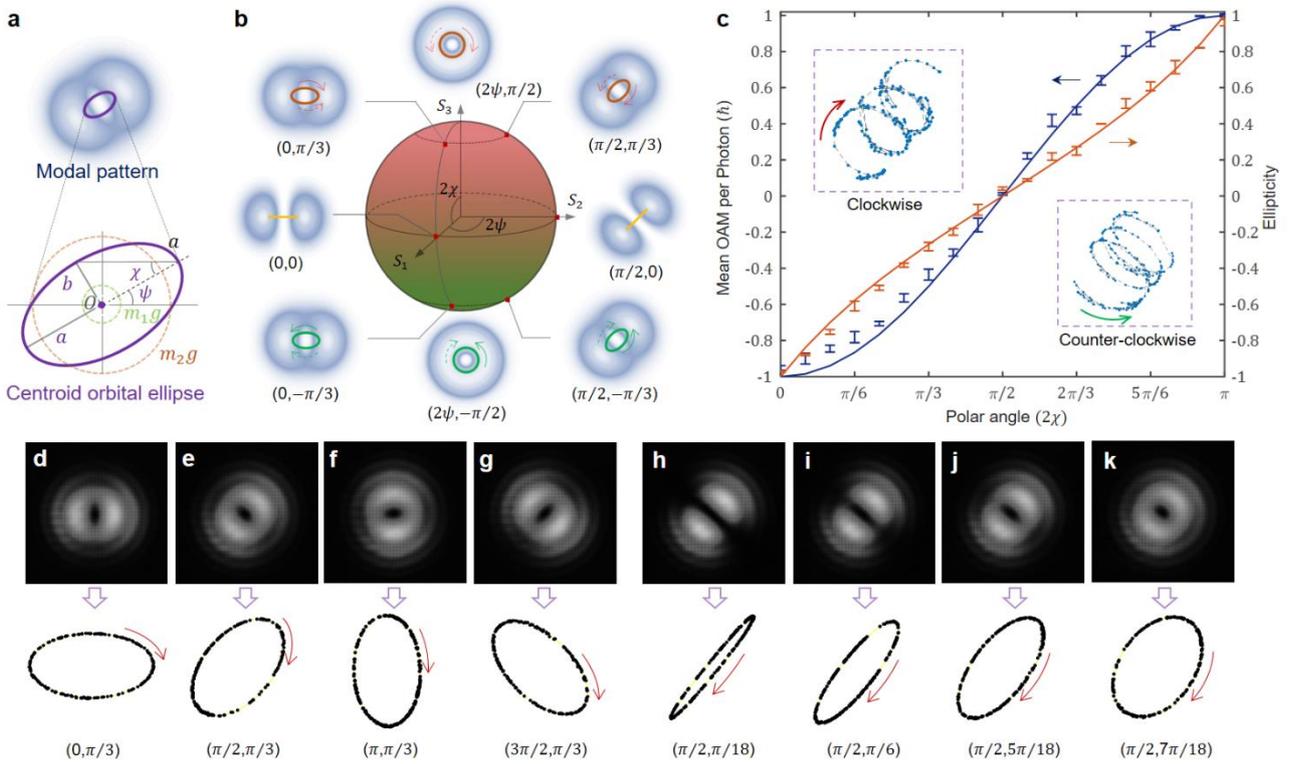

**Figure 4 Geometric characterization of OAM superpositions on modal Poincaré sphere via centroid orbital ellipses**. (**a**) Modal patterns of OAM ($\ell_2 = \pm 1$) superposition state, and their interferometric centroid orbital ellipse generated by dynamical interference with a fundamental LG beam ($\ell_1 = 0$). The inset shows that the centroid ellipse is mathematically derived from the superpositions between two circular centroid orbits (red and green dotted circles). (**b**) The evolution of centroid orbital ellipses is used to characterize arbitrary superposition states on modal Poincaré sphere with OAM eigenbases ($\ell_2 = \pm 1$). (**c**) Measured results of the mean OAM per photon based on the normalized area of ellipses, and the ellipticities ($b/a$) as the ratio of the minor to major axes of ellipses under various polar angles. Note that the insets show different chirality of 3D centroid orbits for OAM superposition states on northern and southern hemispheres. (**d**)-(**g**) Measured modal patterns and their centroid orbital ellipses in four positions along the latitude of $2\chi = \pi/3$. (**h**)-(**k**) Measured modal patterns and the corresponding centroid orbital ellipses in four positions along the longitude of $2\psi = \pi/2$. The dark dots denote the centroid positions extracted from the experimentally collected pictures via centroid algorithm, the yellow lines represent the fitted ellipses, and the arrows near the ellipses indicate the centroid orbiting directions.

From Equations (4) and (5), the major and minor axes of orbital ellipse are given as $a = g(m_1 + m_2)$, $b = g(m_2 - m_1)$, respectively, and the orientation angle of major axis is $\psi = \Delta\beta/2$. From the geometric perspective, the centroid ellipse can be regarded as a result of combination between two perfect centroid circles with different radii, opposite orbiting directions, and a specific phase difference, associated with the superposition states of OAM on modal Poincaré spheres (Figure 4a). The ellipticity can be obtained as $e = \tan\chi = (m_2 - m_1)/(m_1 + m_2)$, associated with the ellipticity angle $\chi$ (see Supplementary Materials). Note that here the minor axis $b$ is defined as a pseudovector, and its sign is determined by the centroid orbiting directions over time. In practice, these signs (or directions) can be distinguished via the phase advance or delay between the x- and y-coordinate values on one time-domain or frequency-phase spectra (see Figures 2b and 2c), rather than by observation via naked eye here (see Supplementary Materials). In addition, the area of orbital ellipse can be given as $S = \pi ab = \pi g^2(m_2^2 - m_1^2)$. Especially, the two OAM eigenbases at the north and south poles have the orbital area of $S = -\pi g^2$ (when $m_1 = 1$, $m_2 = 0$), and $S = \pi g^2$ (when $m_1 = 0$, $m_2 = 1$), respectively. The normalized elliptical area is $S_N = S/S_{max} = m_2^2 - m_1^2 = \sin(2\chi)$, directly associated with the ellipticity angle $2\chi$, which can describe the mean OAM per photon,

$$\bar{L} = (m_2^2 - m_1^2)\ell_2\hbar = \sin(2\chi)\ell\hbar = S_N\ell_2\hbar, \tag{6}$$

where $\hbar = h/2\pi$, and $h$ is the Plank constant (see Supplementary Materials). This provides a novel geometric approach to characterizing abstract optical OAM, particularly its superpositions[46-48].

The OAM superpositions have been fully mapped to the centroid ellipse evolution on modal Poincaré spheres by theoretical derivations above. In the proof-of-concept experiment, we consider the targeted LG beams with the TCNs of $\ell_2 = \pm 1$, and the reference fundamental LG beams ($\ell_1 = 0$). In this case, $\Delta\beta$ only denotes the phase differences between OAM components within the targeted LG beams. The superposition states of OAM were controllably produced by the combination of HWP, QWP, QP, and polarizer (see Figure 2a, Materials and Methods). We demonstrate the experimental results of mean OAM per photon and ellipticity under various polar angles in Figure 4c, and the ellipse evolution of several typical OAM superposition states on modal Poincaré sphere, as shown in Figures 4d-4k. Notably, despite sharing the same ellipses in the mirror positions between the northern and southern hemispheres, the ellipses feature opposite orbiting directions in 3D space over time under the OPDs continuously changed in a unified direction, i.e., a constant $v$ in Equations (4) and (5). This difference is denoted via different color ellipses (solid or dotted arrows), also shown in experimental results (see Figures 4b, S6, and S7, Videos S5 and S6). Note that here the defined parameters of centroid ellipses here share the same characterization for polarization (spin) superpositions on classical Poincaré sphere[49]. It implies that a unified methodology can be established to describe both optical spatial orbital (OAM) and polarization (SAM) superpositions in a 2D Hilbert space using geometric ellipses.

**Centroid orbiting angle subdivision used for linear measurement**
Displacement detection is a fundamental functionality in modern science and technology. The measurement resolution, speed and range are three important performance parameters, but mutually restricted for nearly all types of metrological instruments. In a classical homodyne interferometry, a fundamental trade-off between resolution and dynamic range has always been limited by the nonlinear (sinusoidal) interferometric signals (Figure S9). It is a challenge to achieve a high resolution beyond the wavelength scale that usually requires the wavelength (or phase) subdivision via complex optical or electronic techniques[1,2]. For example, the classical methods of addressing this issue are to adopt quadrature or multiple detection systems, but usually bringing about the low accuracy and complex configureurations[50-54]. Here we demonstrate that the centroid orbiting angles on its circular orbit ($\theta = 0°$ in Figures 1c and 1d) with a linear response to OPDs or phase changes can provide an alternative methodology to break through these limitations (Figure. S9). Actually, the centroid orbiting angles (0~360°) naturally provide a circular subdivision with a linear response, replacing the complex phase interpolation subdivision[1,2,41,42].

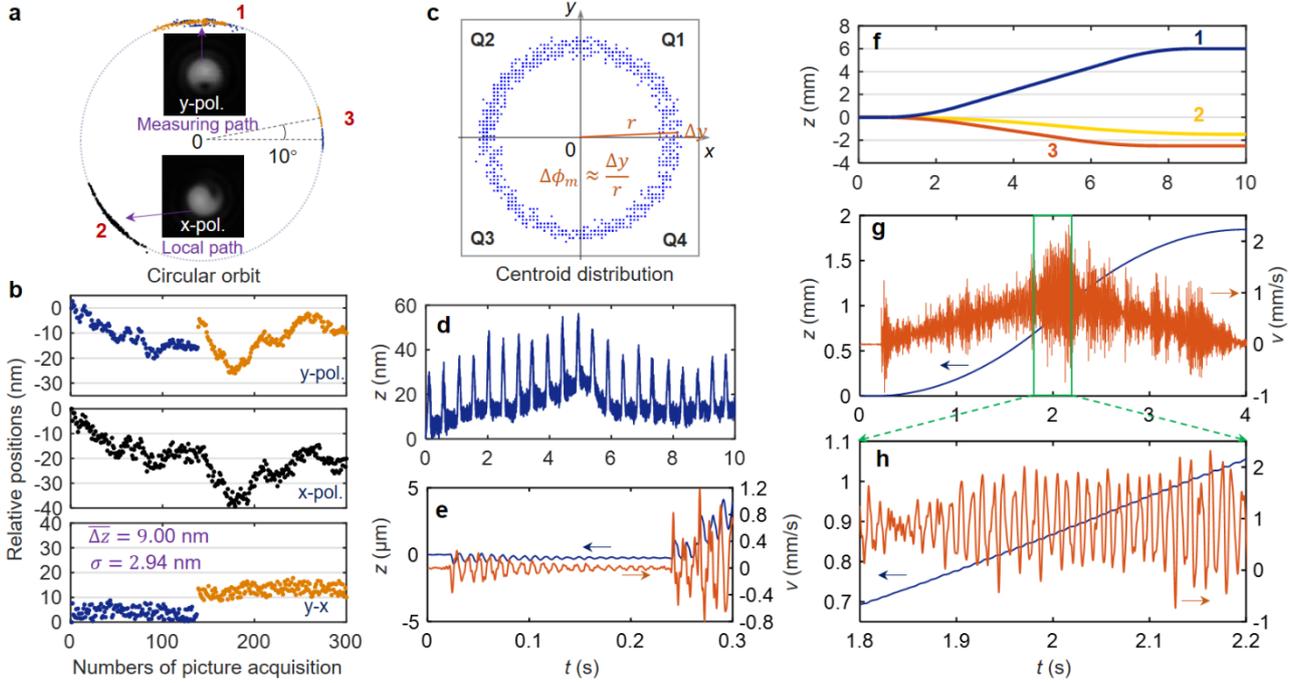

**Figure 5 High-resolution and large-range displacement measurement and monitoring based on centroid orbiting angle detection.** (a) Environmentally robust displacement measurement by synchronously detecting centroid positions of two orthogonally polarized interference patterns via one camera. The local path in x-polarization component (down inset) is used to monitor the environment disturbance to local experimental setup, while the measuring path in y-polarization state (up inset) is to sense the nanoscale displacement driven by a piezoelectric translation stage. (b) Experimental acquisition of a pure interference signal only derived from the moving target by subtracting the signal in the local path from that in the measuring path to eliminate the local environment disturbance. (c) Results of centroids (blue dots) orbiting on their circular orbit monitored by a 4Q photodetector. Fast and high-resolution monitoring results for (d) the nanoscale shaking produced by a puff of consecutive airflow, and (e) the random movement and its instantaneous velocity, produced by a motorized positioning stage when is initiated. (f) Measured results of large-range displacements for three kinds of variable accelerated movements electrically driven by the motorized positioning stage, giving the movements 1 (from 0 to 6 mm), 2 (from 0 to -1.5 mm), and 3 (from 0 to -2.5 mm), respectively. (g) High-resolution monitoring results of linear displacements in space (left y-axis) and meanwhile instantaneous velocities in time (right y-axis) for a variable accelerated movement. Note that the periodical weak shaking superposed to the given movement is caused by the inherent vibration of electrically driven positioning stage, which has been successfully monitored, as shown in (h) with the local amplification.

Since the centroid detection takes advantage of the spatial DoF of light beams, the polarization DoF can be used for synchronous detection using orthogonal polarization states to largely eliminate the environmental disturbance. We first demonstrate this polarization synchronous detection via one camera (see the experimental setup in Figure S14 and the results in Figures 5a and 5b). The centroid groups moved on a circular orbit marked by 1 and 2 denoted in Figure 5a are synchronously extracted by one camera from the measuring and local paths, respectively. The data group 3 is gotten by subtracting the group 2 from the group 1 to eliminate the local environment disturbance (Figure 5b). Here the given displacement is 9 nanometers corresponding to the mean centroid orbiting angle of about 10°. In this experiment, the two coherent light beams with the wavelength of 632.8 nm are $\ell_1 = 0$ and $\ell_2 = 1$. The objective displacement can be generally retrieved as $\Delta z = \Delta\phi/(2k)$, where $\Delta\phi$ is the centroid orbiting angle. More experimental results of the environmentally robust nanoscale linear displacement measurements are exhibited in Figure S10. Note that the standard deviations of about 2 nm in all measured data come from various noises in the measuring system.

In practice, the centroid coordinates can be directly acquired by commercial four-quadrant (4Q) position-sensing photodetectors (Thorlabs PDQ80A, and see Videos S7 and S8). Thanks to the large response bandwidth ($f{\sim}100$ kHz) of these 4Q photodetectors, one can achieve the high-resolution, fast, and large-range movement monitoring. Note that when retrieving the objective displacement, the relative angle between two successive centroid positions should not exceed 180°. Otherwise, this may cause wrong displacement retrievement, because the threshold of judging the signs of centroid orbiting angles is 180° (see Materials and Methods). Accordingly, the maximum instantaneous velocity should be limited to $v_m = \pi f/(2k) = f\lambda/4$ for an effective displacement monitoring by 4Q photodetectors. Here the maximum distinguishable instantaneous velocity is about $v_m = 15.8$ mm/s. The resolution of displacement measurement as another important performance here is associated with an angle resolution estimated as the ratio of the coordinate resolution ($\Delta y$) to the centroid orbital radius ($r$). In our experiment, this subdivision resolution is $\Delta\phi_m = \Delta y/r \approx 2.2°$ (Figure 5c), giving the displacement resolution $\Delta z_m = \Delta y/(2kr) \approx 1.9$ nm. We also demonstrate the elimination of local environment disturbance via two 4Q photodetectors by polarization synchronous detection (see Figure S12 and experimental setup in Figure S14). The primitive experimental results of high-resolution, fast, and large-range monitoring for various linear movements are presented in Figures 5d-5h and S13. Such high resolution of linear displacements in space and meanwhile instantaneous velocities in time attributes to the small subdivision and fast response of centroid orbiting angle detection by the useful 4Q photodetectors. It is yet a big challenge to reach this goal for the conventional detection based on fringe shift or petal rotation by the common used detectors without other assistant techniques. Hopefully, the displacement resolution ruled by $\Delta z_m = \Delta y/(2kr)$ can be further improved by increasing the spatial resolution of detectors (decreasing $\Delta y$) and/or enlarging the centroid orbital radius ($r$). Additionally, the linear range of movement monitoring via the tracking algorithm here actually can surpass 100 millimeters, fundamentally determined by the coherence length of the laser experimentally used.

## Discussion

In this paper, we have shown an interesting centroid orbiting phenomenon of the nearly coaxial interference fields with broken rotational symmetry using OAM light beams. Compared with the conventional interferometry based on fringe shift or petal rotation detection[39-42], the centroid orbiting interferometry by mapping perturbations to centroid positional or orbital information in a 2D plane has two remarkable advantages. One is the centroid ellipses as new detectable DoFs that carry more (elliptic parameters) information, thus allowing for the ultra-sensitive angle distinguishment and characterization of arbitrary OAM superpositions on modal Poincaré spheres. The other is the linear response of centroid orbiting on its circular orbit to the OPD changes that contributes to the high-resolution, fast, and large-range linear movement monitoring directly enabled by the commercial 4Q photodetectors. These interference phenomena and metrological superiorities attribute to the diverse 2D information extracted from the global interference fields via the concept of centroid exploited here, beyond the conventional detection based on fringe shift or petal rotation essentially utilizing the local (or partial) interference fields[41,42].

For the high-resolution and large-range linear movement measurement using 4Q photodetectors, we give the horizontal comparison of its comprehensive performances with the recent vortex interferometry (see Tab. S1). Furthermore, we also exhibit the radar map for further comparison among several typical interferometric schemes in terms of measurement resolution, range, velocity, cost, and complexity (see Figure S15). From overall comparisons, the centroid orbiting interferometry here used for linear measurement has good compositive performances, especially possessing the advantages of simple technique and low cost. Despite not showing a higher resolution (smaller than 1 nm) than other schemes at present, we believe that such performance can be further improved by increasing centroid orbital radii and detectable spatial resolution, especially combining with other subdivision and assistant techniques.

In conclusion, the centroid orbiting metrology proposed here by fully exploiting the spatial orbital DoF of structured light can provide more detectable 2D positional and geometric information compared with the conventional interferometry. We have experimentally demonstrated its potentials to multifunctional and nanoscale metrological applications. The results show that it can break though the conventional interferometric limitations, such as the detectable information shortage and short measuring range. We expect that this new interferometric paradigm may pave the way to develop more advanced laser interferometric technologies, and even offer an inspiration in the subdisciplines linking optics with geometry or astronomy.

## Materials and Methods

**Experimentally generating perfect LG Gaussian beams with circular or donut shapes**
In the experiments, the perfect light beam with a circular shape must be guaranteed for all the measurements based on centroid orbiting angle detection here. The light source emitted from a helium-neon laser in the experiments was first filtered by passing through a single-mode fiber (SMF) to generate a perfect Gaussian beam (see Figures S5, S8, and S14). When experimentally producing perfect higher-order LG beams carrying OAM through a QP or SPP, the center axis of the circular Gaussian beam must be aligned with the centers of these plates.

**Experimentally producing the tiny inclined angles between two interference light beams**
In the experiment, the gradually variable inclined angles were produced by controlling the radio of small transverse displacement ($\Delta x$, about 5.0 μm per step) of the mirror to the longitudinal length ($L \approx 1150.0$ mm) for the probing light path via a six-dimensional alignment jig (see Supplementary Figure. S5). In this case, the inclined angle step can be obtained as $\theta \approx \Delta x/L = 0.00025°$. The dynamic interference patterns were produced by manually changing the OPDs by pressing the bracket where the beam combiner (BS2), and collected by a high-speed camera (see Figure S5).

**Extracting centroid coordinates and plotting centroid orbits from collected pictures**

The centroid coordinates were calculated via centroid algorithm for interference patterns experimentally collected by a camera. The gray values $P(i,j)$ of each pixel of interference patterns can be gotten from the collected pictures, where $i$ and $j$ denote the $i$-th row and $j$-th column. Since the gray matrix of pictures are usually in direct proportion to the power densities of collected light, the centroid coordinate $(\bar{x}, \bar{y})$ of the interference patterns in our experiment can be calculated as $\bar{x} = \sum_{j=1}^{n} j \sum_{i=1}^{m} P(i,j) / \left[ \sum_{i=1}^{m} \sum_{j=1}^{n} P(i,j) \right]$ and $\bar{y} = \sum_{i=1}^{m} i \sum_{j=1}^{n} P(i,j) / \left[ \sum_{i=1}^{m} \sum_{j=1}^{n} P(i,j) \right]$, where $m$ and $n$ are the numbers of total rows and columns of the pictures, respectively. The centroid orbital ellipses and their elliptic parameters including major and minor axes, orientation angle of major axis, and elliptical center position were obtained from the discrete centroid points via fitting algorithm.

**Experimentally generating arbitrary OAM superposed states on modal Poincaré spheres**

In this experiment, we used the combination of HWP, QWP, QP, and polarizer to generate and control the arbitrary superposition states of OAM beams with $\ell_2 = \pm 1$ on modal Poincaré spheres (see Figures 3 and S8). The generation of arbitrary superposition states can be expressed by means of matrix multiplication,

$$\mathbf{E} = \mathbf{M}_p \cdot \mathbf{M}_v \cdot \mathbf{M}_{\lambda/4} \cdot \mathbf{M}_{\lambda/2} \cdot \mathbf{E}_{in}, \tag{7}$$

where the Jones matrixes of HWP, QWP, QP[55,56], and polarizer are respectively given as

$$\mathbf{M}_{\lambda/2} = \begin{bmatrix} \cos(2\theta_1) & \sin(2\theta_1) \\ \sin(2\theta_1) & -\cos(2\theta_1) \end{bmatrix}, \tag{8}$$

$$\mathbf{M}_{\lambda/4} = \begin{bmatrix} 1 - i\cos(2\theta_2) & -i\sin(2\theta_2) \\ -i\sin(2\theta_2) & 1 + i\cos(2\theta_2) \end{bmatrix}, \tag{9}$$

$$\mathbf{M}_v = \begin{bmatrix} \cos\phi & \sin\phi \\ \sin\phi & -\cos\phi \end{bmatrix}, \tag{10}$$

and

$$\mathbf{M}_p = \begin{bmatrix} \cos^2\theta_3 & \frac{1}{2}\sin(2\theta_3) \\ \frac{1}{2}\sin(2\theta_3) & \sin^2\theta_3 \end{bmatrix}. \tag{11}$$

For convenience, here we set the light input and output both with x-polarization, i.e., $\mathbf{E}_{in} = [1 \ \ 0]^T$, and $\theta_3 = 0$, thus, the arbitrary superposition states on modal Poincaré spheres can be generated as

$$E_x = \mathbf{M}_p \cdot \mathbf{M}_v \cdot \mathbf{M}_{\lambda/4} \cdot \mathbf{M}_{\lambda/2} \cdot \mathbf{E}_{in} = \cos(2\theta_1 - \phi) - i\cos[2(\theta_2 - \theta_1) - \phi]$$

$$\propto [\cos(\theta_2 - 2\theta_1) + \sin(\theta_2 - 2\theta_1)]e^{-i(\phi - \theta_2)} + i[\sin(\theta_2 - 2\theta_1) - \cos(\theta_2 - 2\theta_1)]e^{i(\phi - \theta_2)}$$

$$\propto \cos\left(\theta_2 - 2\theta_1 - \frac{\pi}{4}\right) e^{-i\left(\phi - \theta_2 + \frac{\pi}{4}\right)} + \sin\left(\theta_2 - 2\theta_1 - \frac{\pi}{4}\right) e^{i\left(\phi - \theta_2 + \frac{\pi}{4}\right)}. \tag{12}$$

It shows that the longitude ($2\psi$) and latitude ($2\chi$) angles of modal states on Poincaré spheres can be obtained by rotating the work axes ($\theta_1$ and $\theta_2$) of HWP and QWP, given by $2\psi = 2\theta_2 - \frac{\pi}{2}$ and $2\chi = 2\theta_2 - 4\theta_1 - \frac{\pi}{2}$, respectively.

**Principle and algorithm of movement monitoring based on centroid orbiting angles**

The azimuthal angle of centroid orbiting position P on its circular orbit can be obtained by $\phi = \text{atan2} = (\bar{y}/\bar{x})$, where $(\bar{x}, \bar{y})$ denotes the coordinate in Cartesian coordinates obtained by 4Q photodetectors. When constraining the angle range from $-\pi$ to $\pi$, the relative azimuthal angle between two centroid positions $P_{i-1}$ $(\bar{x}_{i-1}, \bar{y}_{i-1})$ and $P_i$ $(\bar{x}_i, \bar{y}_i)$ can be judged as

$$\Delta\phi_i = \begin{cases} \phi_i - \phi_{i-1}, & (-\pi \ll \phi_i - \phi_{i-1} \ll \pi) \\ \phi_i - \phi_{i-1} - 2\pi, & (\phi_i - \phi_{i-1} > \pi), \\ \phi_i - \phi_{i-1} + 2\pi, & (\phi_i - \phi_{i-1} < -\pi) \end{cases} \qquad (13)$$

where $i = 2,3,4 \ldots$. These two successive centroid positions correspond to a displacement step of $\Delta z_i = \Delta\phi_i/(2k)$. Therefore, an arbitrary complex linear movement can be real-time monitored (see Figure S13), and the instantaneous velocity can be deduced as $v_i = \Delta z_i/\Delta t = \Delta\phi_i/(2k\Delta t)$. Note that the fixed time interval ($\Delta t$) is given by the response bandwidth ($f$) of 4Q photodetectors, i.e., $\Delta t = 1/f$. When considering the constraint range of $\Delta\phi_i \in [-\pi, \pi]$, it determines the maximum distinguishable instantaneous velocity for the moving target, $v_m = \pi f/(2k) = f\lambda/4$.

**Acknowledgements**

We would like to thank Shuangchun Wen from School of Physics and Electronics, Hunan University for providing the laboratory and optical platform. This work was supported by the Natural Science Foundation of China (NSFC) (62275092, 61905081), and the Fundamental Research Funds for the Central Universities.

**Author contributions**

L.F. conceived the concept and experiments, performed the theoretical derivation and numerical simulations. L.F. and J.C. performed the experimental measurement and data analyses. L.F. wrote the manuscript and supervised the project. All authors participated in discussions and contributed to the editing of the article.

**Conflict of interest**

The authors declare no competing interests.

**Supplementary information**

Including Supplementary Text, Supplementary part I-III, Figures S1-S15, References (S1-S10), and Videos S1-S8.

**References**

1. P. J de Groot, A review of selected topics in interferometric optical metrology. *Rep. Prog. Phys.* **2018**, 82, 056101.

2. J. Watchi, S. Cooper, B. Ding, C. M. Mow-Lowry, C. Collette, Contributed Review: A review of compact interferometers. *Rev. Sci. Instrum.* **2018**, 89, 121501.
3. A. A. Michelson, E. W. Morley, On the relative motion of the earth and the luminiferous ether. *Am. J. Sci.* **1887**, 34, 333.
4. B. P. Abbott, et al. Observation of gravitational waves from a binary black hole merger. *Phys. Rev. Lett.* **2016**, 116, 061102.
5. G. Sagnac, The demonstration of the luminiferous aether by an interferometer in uniform rotation. *Comptes Rendus* **1913**, 157, 708.
6. D. Malacara, Twyman-Green interferometer in Optical Shop Testing (Wiley & Sons, New York, 1992).
7. Z. Cheng, H. Gao, Z. Zhang, H. Huang, J. Zhu, Study of a dual-frequency laser interferometer with unique optical subdivision techniques. *Appl. Opt.* **2006**, 45, 2246.
8. R. Teti, K. Jemielniak, G. O'Donnell, D. Dornfeld, Advanced monitoring of machining operation. *CIRP Annu. Manuf. Technol*. **2010**, 59, 717.
9. M. V. Mantravadi, D. Malacara, Newton, Fizeau, & Haidinger interferometers Optical Shop Testing ed D. Malacara (New York: Wiley) pp 361-94 (2007).
10. C. Collette, et al. Inertial Sensors for Low-Frequency Seismic Vibration Measurement. *Bull. Seismol. Soc. Am.* **2012**, 102, 1289.
11. M. K. Zhou, et al. Note: A three-dimension active vibration isolator for precision atom gravimeters. *Rev. Sci. Instrum.* **2015**, 86, 046108.
12. L. Allen, M. W. Beijersbergen, R. J. C. Spreeuw, J. P. Woerdman, Orbital angular momentum of light and the transformation of Laguerre-Gaussian laser modes. *Phys. Rev. A* **1992**, 45, 8185.
13. A. M. Yao, M. J. Padgett, Orbital angular momentum: origins, behavior and applications. *Adv. Opt. Photon.* **2011**, 3: 161.
14. G. Milione, H. Sztul, D. A. Nolan, R. R. Alfano, Higher-order Poincaré sphere, Stokes parameters, and the angular momentum of light. *Phys. Rev. Lett.* **2011**, 107, 053601.
15. D. Naidoo, et al. Controlled generation of higher-order Poincaré sphere beams from a laser. *Nat. Photon.* **2016**, 10, 327.
16. Q. Zhang, Cylindrical vector beams from mathematical concepts to applications. *Adv. Opt. Photon.* **2009**, 1, 1.
17. A. Forbes, M. de, Oliveira, M .R. Dennis, Structured light. *Nat. Photon.* **2021**, 15, 253.
18. C. He, Y. Shen, A. Forbes, Towards higher-dimensional structured light, *Light Sci. Appl.* **2022**, 11:205.
19. Z. Wan, H. Wang, Q. Liu, X. Fu, Y. Shen, Ultra-Degree-of-Freedom Structured Light for Ultracapacity Information Carriers, *ACS Photonics* **2023**, 10, 7, 2149.
20. Y. Shen, Q. Zhang, P. Shi, L. Du, X. Yuan, A. V. Zayats, Optical skyrmions and other topological quasiparticles of light. *Nat. Photon.* **2024**, 18, 15.
21. L. Fang, J. Wang, Emerging optical spin and chirality by three-dimensional evanescent field coupling, Phys. Rev. A **2024**, 110, 063514.
22. R. Dorn, S. Quabis, G. Leuchs, Sharper focus for a radially polarized light beam. P*hys. Rev. Lett.* **2003**, 91, 233901.
23. M. P. J. Lavery, F. C. Speirits, S. M. Barnett, M. J. Padgett, Detection of a Spinning Object Using Light's Orbital Angular Momentum. *Science* **2013**, 341, 537.

24. L. Fang, M. J. Padgett, J. Wang, Sharing a common origin between the rotational and linear Doppler effects. *Laser & Photon. Rev.* **2017**, 11, 1700183.
25. A. Aiello, P. Banzer, M. Neugebauer, G. Leuchs, From Transverse Angular Momentum to Photonic Wheels. *Nat. Photonics* **2015**, 9, 789.
26. K. Y. Bliokh, F. Nori, Transverse and longitudinal angular momenta of light. *Phys. Rep.* **2015,** 592, 1.
27. K. I. Willig, S. O. Rizzoli, V. Westphal, R. Jahn, S. W. Hell, STED microscopy reveals that synaptotagmin remains clustered after synaptic vesicle exocytosis. *Nature* **2006**, 440, 935.
28. N. Bozinovic, et al. Terabit-scale orbital angular momentum mode division multiplexing in fibers. *Science* **2013**, 340, 1545.
29. M. Meier, V. Romano, T. Feurer, Material processing with pulsed radially and azimuthally polarized laser radiation. *Appl. Phys. A* **2007**, 86, 329.
30. M. J. Padgett, R. Bowman, Tweezers with a twist. *Nat. Photonics* **2011**, 5, 343.
31. V. Shvedov, A. R. Davoyan, C. Hnatovsky, N. Engheta, W. Krolikowski, A long-range polarization-controlled optical tractor beam. *Nat. Photonics* **2014**, 8, 846.
32. A. P. Greenberg, G. Prabhakar, S. Ramachandran, High resolution spectral metrology leveraging topologically enhanced optical activity in fibers. *Nat. Commun.* **2020**, 11, 5257.
33. L. Fang, Z. Wan, A. Forbes, J. Wang, Vectorial Doppler metrology. *Nat. Commun.* **2021**, 12: 4186.
34. L. Kong, et al. High capacity topological coding based on nested vortex knots and links. *Nat. Commun.* **2022**, 13, 2705.
35. S. Fürhapter, A. Jesacher, S. Bernet, M. Ritsch-Marte, Spiral interferometry. *Opt. Lett*. **2022**, 30, 1953.
36. V. D'Ambrosio, et al., Photonic polarization gears for ultra-sensitive angular measurements, *Nat. Commun. 2013*, 4:2432.
37. R. Barboza, et al., Ultra-sensitive measurement of transverse displacements with linear photonic gears, *Nat. Commun. 2022*, 13:1080.
38. H. Zang, et al., Ultrasensitive and long-range transverse displacement metrology with polarization-encoded metasurface, *Sci. Adv.* **2022**, 8, eadd1973.
39. X. Hu, B. Zhou, Z. Zhu, W. Gao, C. Rosales-guzmán, In situ detection of a cooperative target's longitudinal and angular speed using structured light. *Opt. Lett***. 2019***, 44, 3070.
40. G. Verma, G. Yadav, Compact picometer-scale interferometer using twisted light. *Opt. Lett.* **2019,** 44, 3594.
41. G. Ye, T. Yuan, Y. Zhang, T. Wang, X. Zhang, Recent progress on laser interferometry based on vortex beams: Status, challenges, and perspectives. *Opt. Laser Eng.* **2024**, 172, 107871.
42. J. T. Dong, E. X. Zhao, L. Y. Xie, Y. Y. Li, Z. P. Tian, X. L. Xie, Optical vortex interferometer: An overview of interferogram demodulation methods for dynamic phase measurement. *Opt. Laser Eng.* **2024, 175,** 108044.
43. J. Leach, S. Keen, M. J. Padgett, Direct measurement of the skew angle of the Poynting vector in a helically phased beam. *Opt. Express* **2006**, 14, 11919.
44. M. J. Padgett, J. Courtial, Poincaré-sphere equivalent for light beams containing orbital angular momentum, *Opt. Lett.* **1999**, 25, 430.
45. T. Malhotra, et al., Measuring Geometric Phase without Interferometry, *Phys. Rev. Lett.* **2018**, 120, 233602.

46. G. Molina-Terriza, J. P. Torres, L. Torner, Management of the angular momentum of light: preparation of photons in multidimensional vector states of angular momentum. *Phys. Rev. Lett.* **2002,** 88, 013601.
47. L. Fang, J. Wang, Optical angular momentum derivation and evolution from vector field superposition. *Opt. Express* **2017**, 25, 23364.
48. J. Cheng, L. Fang, J. Chen, Y. Zhou, F. Fan, L. Miao, C. Zhao, Optical centroid ellipses beyond polarization ellipses, *Opt. Lett.* **2024**, 50, 97.
49. D. Goldstein, Polarized light (3nd ed.), Marcel Dekker Inc., New York (2011).
50. P. L. M. Heydemann, Determination and correction of quadrature fringe measurement errors in interferometers. *Appl. Opt.* **1981**, 20, 3382.
51. P. Gregorčič, T. Požar, J. Možina, Quadrature phase-shift error analysis using a homodyne laser interferometer. *Opt. Express* **2009**, **17**, 16322.
52. M. J. Downs, K. W. Raine, An unmodulated bi-directional fringe-counting interferometer system for measuring displacement. *Precis. Eng.* **1979 1**: 85.
53. A. Dorsey, R. J. Hocken, M. A. Horowitz, low cost laser interferometer system for machine tool applications. *Precis. Eng.* **1983**, M5: 29.
54. P. de Groot, Homodyne interferometric receiver and calibration method having improved accuracy and functionality US Patent 5,663,793 (1997).
55. L. Marrucci, C. Manzo, D. Paparo, Optical Spin-to-Orbital Angular Momentum Conversion in Inhomogeneous Anisotropic Media. *Phys. Rev. Lett.* **2006**, **96**: 163905.
56. Y. Liu, X. Ling, X. Yi, *et al*. Realization of polarization evolution on higher-order Poincare sphere with metasurface, *Appl. Phys. Lett.* **2014,** 104: 191110.